\documentclass[preprint]{aa}
 \input psfig.tex
 \usepackage{graphicx}
 \usepackage{natbib}
 \usepackage{array}
\bibpunct{(}{)}{;}{a}{}{,}
 \usepackage{rotating}
 \usepackage{latexsym}
 \usepackage{amssymb}
\usepackage{amsmath}
\usepackage{graphics}
\usepackage{fancyhdr}
\newcommand{\de}{{\rm d}}
\newcommand{\bea}{\begin{eqnarray}}
\newcommand{\eea}{\end{eqnarray}}
\newcommand{\f}{\frac}
\begin{document}

\title{Negative feedback effects on star formation history and cosmic reionization}
\subtitle{}
     \author{ Lei Wang \inst{1}, Jirong Mao \inst{2}, Shouping Xiang \inst{1} \&
 Ye-Fei Yuan \inst{1}            }
     \offprints{Lei Wang }
     \institute{$^1 $ Center for Astrophysics, University of Science and Technology
of China, Hefei, 230026, China  \\
  $^2 $ Yunnan Observatory, National Astronomical Observatories, Chinese Academy of
Sciences, P.O.Box 110, Kunming, Yunnan Province, 650011, China.
                \email{wsl2008@mail.ustc.edu.cn; jirongmao@ynao.ac.cn; spxiang@ustc.edu.cn;
yfyuan@ustc.edu.cn}}

\abstract {The mechanical and radiative feedback that exists in the star
 formation history affects the subsequent star formation rate. }
 {After considering the effects of negative feedback on the process
 of star formation, we explore the relationship between star formation
 process and the associated feedback, by investigating how the mechanical
 feedback from supernovae(SNe) and radiative feedback from luminous objects
 regulate the star formation rate and therefore affect the cosmic reionization.} 
{Based on our present knowledge of the negative feedback theory and some
 numerical simulations, we construct an analytic model in the framework
 of the Lambda cold dark matter model. In certain parameter regions, our
 model can explain some observational results properly.}
 { In large halos($T_{\rm vir}>10^4K$), both mechanical and radiative
 feedback have a similar behavior: the relative strength of negative
 feedback reduces as the redshift decreases. In contrast, in small halos
($T_{\rm vir}<10^4K$) that are thought to breed the first stars at early time,
 the radiative feedback gets stronger when the redshift decreases. And the star
 formation rate in these small halos depends very weakly on the
 star-formation efficiency.}
 {Our results show that the radiative feedback is important
 for the early generation stars. It can suppress the star formation
 rate considerably. But the mechanical feedback from the SNe explosions is
 not able to affect the early star formation significantly. The early star
 formation in small-halo objects is likely to be self-regulated. The radiative
 and mechanical feedback dominates the star formation rate of the PopII/I
 stars all along. The feedback from first generation stars is very strong and
 should not be neglected. However, their effects on the cosmic reionization are
 not significant, which results in a small contribution to the optical depth
 of Thomson scattering.}
\titlerunning{Feedback, SFH, and cosmic reionization}
\authorrunning{Wang, et al.}
\maketitle

\section{Introduction}

In the past decades, the so-called ``bottom-up" hierarchical scenario
 for the large-scale structure formation in the Lambda cold dark matter
 ($\Lambda$CDM) cosmogonies has been getting decisive support from
 more and more observations (e.g. from HST, WMAP, SDSS) and from
 high-resolution N-body/hydrodynamics numerical simulations. Owing to the
 complexity of the baryonic evolution in the radiative background and
 gravitational field, the galaxy formation and evolution in dark matter
 halos and the reionization history of IGM have not been fully understood
 yet.
%
The difficulties stem from our lack of knowledge of those early formed
 objects that are unobservable at present. The formation of the
 first objects is relevant to the cosmic reionization history.
 Recent detections of the Gunn-Peterson trough\cite{gunn1965}
 in the spectra of QSOs with $z\simeq 6$ indicate less 50\% neutral hydrogen at
 $z \sim 6.5$ \cite{fan,loeb}. The ongoing observations by the WMAP satellite of
 cosmic microwave background (CMB) and the highest redshift QSOs put very
 tight constraints on the reionization history of the Universe. The WMAP
 five-year observation manifests the Thomson scattering optical depth,
 $\tau_{\rm e} = 0.084^{+0.016}_{-0.016}$ \cite{Komatsu08}, which
 suggests that our Universe might be reionized during the period of redshift
 $9.4 \le z_{\rm re} \le 12.2$.

It is well-known that chemical elements heavier than lithium are produced
 exclusively through stellar nucleosynthesis. Some of the first generation
 stars (the so-called population III stars, hereafter, PopIII) die as SNe
 explosions, which can expel heavier elements into the intergalactic medium (IGM).
 When the metal elements in IGM are enriched to a certain threshold $Z_{\rm crit}$,
 the population II/I stars (hereafter PopII/I) will form and take the place of
 the first stars to light the universe. Thence,
 the SNe from PopIII stars determine the transition from PopIII to
 PopII/I. The existence of PopIII stars can help in explaining the metal
 enrichment from $Z\simeq 10^{-12}-10^{-10}$ to the lowest metallicity of PopII stars
 $Z\simeq 10^{-4}-10^{-3}$, the formation of massive black holes, the reionization
 of the universe, the starting engine for the formation of the first galaxies
 and the G-dwarf, and so on\cite{cf05}. But there is still no hope in observing
 the first generation stars until the launch of the
 $James\ Webb\ Space\ Telescope$ (JWST), the successor of the
 $Hubble\ Space\ Telescope$ (HST) \cite{bl01}. JWST is expected to find the
 pair instability SNe (PISNe) from massive PopIII stars \cite{wise05}.

 Due to the absence of the observational data of
 very high-redshift ($z\ge 10$) objects, theoretical investigations are mainly
 based on numerical simulations. Some works concentrated on the effects of 
 the first-generation SNe explosions\cite{Yoshida03,KY05,Greif07}, while 
 the others focused on the strong stellar
 and galactic winds from PopIII stars\cite{MEM06,Ricotti08}. Based on these
 works, we put forward a model to describe the global effects from the PopIII
 stars at an early time, such as the SFR density, the IGM reionization, and so on.

In this paper, we study the effects of the mechanical feedback from SNe and
 the radiative feedback from stars and UV background, especially the negative 
 feedback effects on the SFH and the cosmic reionization history.
 The radiative feedback from PopIII stars includes the ISM photoevaporation 
 and IGM reionization.
 The outline of this paper is as follows. In section 2, we describe the 
 evolution of dark matter halos in the $\Lambda$CDM model.
 Within this framework, the SFR in a halo can be expressed as an analytic 
 formula with some free parameters.
 Mechanical feedback from stars is studied in section 3. In section 4, we
 introduce an analytic model to deal with the radiative feedback from PopIII.
 Finally, our discussion and conclusions are presented in the last section.
 Throughout this paper, we adopt the cosmological parameters consistent 
 with the 5 years WMAP data:$\Omega_{\rm m}h^{\rm 2} = 0.1358$, 
 $\Omega_\Lambda = 0.726$, $\Omega_{\rm b}=0.0456$, $h = 0.7$, 
 $\sigma_8=0.812$\cite{Komatsu08}.

\section{Cold dark matter model and star formation in galaxies}
\subsection{Redshift evolution of dark matter halos}
In the hierarchical clustering scenario of CDM halos, a simple and successful
 model for the formation and distribution of spheroidal or ellipsoidal collapsing
 objects was presented in last century based on the theory of Gaussian random
 fields about cosmological density perturbation \cite{ps74,BBKS86}. In this
 scenario, gravity governs almost the whole process. The dark matter halos
 increase their own mass through accreting matter and merging with each
 other\cite{LC93}. Based on this model, Sasaki (1994) proposed an analytic
 formula to describe the formation 
 and evolution of the CDM halos.
 In this formalism, the number density of collapsed objects with mass in the
 range $(M, M + \de M)$, which are formed at the redshift interval
 $(z_{\rm c}, z_{\rm c} + \de z_{\rm c})$ and survive till redshift $z$,
 is \cite{sasaki94,chiu},

\bea
N(M,z,z_{\rm c})\de M\de z_{\rm c} 
= {\alpha} N_M(z_{\rm c})\left(\f{\delta_c}{D(z_{\rm c})
\sigma(M)}\right)^2 \f{\dot{D}(z_{\rm c})}{D(z)} \nonumber\\
\times \f{\de z_{\rm c}}{H(z_{\rm c}) (1 + z_{\rm c})}\de M.
\label{eqnmPS}
\eea
Here the overdot represents the time derivative, and $N_M(z_{\rm c})~ \de M$
 is the number of collapsed objects per unit comoving volume within a mass range
 $(M, M+\de M)$ at redshift $z_{\rm c}$ \cite{ps74}. In Eq.~(\ref{eqnmPS}),
 $\delta_c$ is a constant, usually taken to be
 1.686 in a matter-dominated flat universe $(\Omega_{\rm m}=1)$. This
 value is quite insensitive to the cosmological parameters \cite{ecf96}, where 
 $H(z)$ is the Hubble parameter,
 $D(z)$ the growth factor for linear perturbations, and
 $\sigma(M)$ the rms mass fluctuation on a mass scale~$M$. In addition,
 $N(M,z,z_{\rm c})$ in Eq.~(\ref{eqnmPS}) represents the formation rate of halos
 weighted by their survival probability. For more accuracy, here we use
 Sheth \& Tormen (1999)'s modified formula for the expression of
 $N_M(z_{\rm c})\de M$, which fits the numerical simulations
 better than the original one, especially at high redshift. Numerical
 simulations also indicate $\alpha= 0.707$
 (see Sheth, Mo \& Tormen 2001 for more details).

\subsection{Star-formation rate in galaxies and the cosmic SFR density}

Stars or even galaxies are confirmed to form through cooling and condensation
 of baryons within the DM halos \cite{White78}. The formation and evolution
 of galaxies and the associated star formation histories have been studied
 extensively via both numerical simulations and semi-analytic
 models \cite{cen92,chiu,CS02,springel03,nagamie06}.
 Following Cen \& Ostriker (1992), Chiu \& Ostriker (2000) and Choudhury \&
 Srianand (2002), we assume\footnote{In fact, this is not the most fundamental
 assumption(see $Appendix\ A$ for details).} that the SFR in a halo with
 mass $M$ at $z$ that has collapsed at an earlier redshift $z_{\rm c}$, 
 is given by
\bea
\dot{M}_{\rm SF}(M,z,z_{\rm c}) = f_*M\f{\Omega_{\rm b}}{\Omega_{\rm m}}\times
\f{t(z)-t(z_{\rm c})}{\kappa^2t_{\rm dyn}^2(z_{\rm c})}{\exp}\left(\f{t(z)-t(z_{\rm c})}
{-\kappa t_{\rm dyn}(z_{\rm c})}\right).
\label{SFR}
\eea
Here, $f_*$, $t(z)$, $t_{\rm dyn}$  are the fraction of total baryonic mass 
in a halo that will be converted to stars, the age of the Universe at
redshift $z$, and the dynamical time scale, respectively.  In our model, we 
take $f_*^{\rm III}=0.07$\cite{CF06} for PopIII stars and $f_*^{\rm II}=0.3$ 
for PopII/I stars. The dynamical time scale $t_{\rm dyn}$ is given by 
(Chiu \& Ostriker 2000),
\bea
t_{\rm dyn}(z) &=& \sqrt{\f{3\pi}{32 G \rho _{\rm vir}(z)}},
\label{tdyn}
\eea
and here,
\bea
\rho_{\rm vir}(z) &=& \Delta_c(z) \rho_c(z), \nonumber \\
\Delta_c (z) &=& 18 \pi ^2 + 82 d(z) - 39 d^2(z), \nonumber\\
d(z) &=& \f{\Omega_m ( 1 + z )^3}{\Omega_m ( 1 + z )^3 + \Omega_{\Lambda}} - 1, \nonumber \\
\rho_c(z) &=& \f{3 H^2(z)}{8 \pi G}. \nonumber 
\eea
 The duration of star formation activity in a halo depends on the value of
 $\kappa $. Note that $\kappa \to 0$ corresponds to the star formation
 occurring in a single burst. By using a 3D cosmological code that includes
 most of the needed physics to simulate the formation and evolution of the first
 galaxies, Ricotti et al. (2002a;2002b;2008) show that the first luminous
 objects (``small-halo objects") are characterized by ``bursting" star formation. 
 In this paper, we thus set $\kappa_{\rm III}=0.01$ for PopIII SF\cite{Samui07}
 and $\kappa_{\rm II}=1$ for PopII/I\cite{cen92,chiu,CS02}.

Consequently, the cosmic SFR per unit comoving volume (SFR density, hereafter SFRD)
 at a redshift $z$ can be written as
\begin{equation}
\dot{\rho}_{\rm SF}(z)=\int\limits_z^{\infty}\de z_{\rm c}
\int\limits_{M_{\rm low}}^{\infty}N(M',z,z_{\rm c})
\times \dot{M}_{\rm SF}(M',z,z_{\rm c})\de M',
\label{eqnsfr}
\end{equation}
where the lower mass cutoff $M_{\rm low}(z)$ at a given epoch depends on the
 cooling efficiency of the gas and the different feedback processes. Following
 Barkana $\&$ Loeb (2005), the lower mass cutoff can be expressed as
\begin{equation}
 M_{\rm low}(z) = 9.2 \times 10^7
M_{\odot} \left( \frac{V_{\rm c}}{16.5{\rm km\ s^{-1}}} \right)^3 \left( \frac{1+z}{10}
\right)^{-\frac{3} {2}},
\label{eqM_low}
\end{equation}
 where the circular velocity 
$V_{\rm c}=(2k_{\rm B}T_{\rm vir}/\mu m_{\rm p})^{1/2}$,
the mean molecular weight $\mu=0.61$, and $T_{\rm vir}$ is the virial
 temperature of halos. Halos
 with $T_{\rm vir}\geq 10^4 K$ are able to cool via atomic transitions. It is
 usually considered as standard in most semi-analytic models that the minimum
 mass of star-forming halos is $M_{\rm low}(z=6) \sim 10^8M_{\odot}$.
 When the H$_2$ content of the gas is considerable, molecular line cooling
 can make gas condense within the small halos ($T_{\rm vir}\sim 10^3 K$) and
 eventually form stars \cite{tsrbap97,Haiman00,bl05,schneider06,Samui07}.
 In the $\Lambda$CDM model, the number of halos with masses $10^5-10^8M_{\odot}$
($T_{\rm vir}< 10^4 K$) that are expected to possess PopIII stars at
 high-redshift decreases after $z\sim 10$ due to the DM halos merging
 process \cite{Mo02}, so these small-halo objects can dominate the galaxy mass
 function until at least redshift $z\sim 10$ \cite{Ricotti02b}.
 Finally, we consider the atomic cooling model ($T_{\rm vir}\sim 10^4 K$) for
 PopII/I and molecular cooling model ($T_{\rm vir}\sim 10^3 K$) for PopIII.
 It is notable that the $M_{\rm low}$ mentioned above corresponds to the neutral
 regions only.

On the other hand, the Lyman-Werner (912-1108\AA) background can also inhibit the
 star formation \cite{Haiman97,Ricotti02a,Yoshida03}. However, it is interesting
 that the positive feedback of H$_2$ re-formation
 (e.g. in front of HII regions and inside relic HII regions) can counterbalance
 the negative feedback from LW background \cite{Ricotti01,Ricotti02a}.
 As a consequence, the formation of small-mass galaxies is not suppressed,
 so in this paper we do not consider the LW background feedback by 
 assuming that positive feedback dominates and star formation in small 
 galaxies is not suppressed.

\subsection{PopIII mass fraction}
With the evolution of the Universe, the characteristics of star formation 
changes from a metal-free, massive-star-dominated
 (PopIII) mode to a metal-enriched, normal-star-dominated (PopII/I) mode.
 Numerical simulations indicate that this transition occurs when the metal has
 been enriched to a critical value,
 $Z_{\rm crit}\simeq 10^{-3.5\sim -4}Z_\odot$ \cite{BYH03,SS07}.
 If most of the first generation stars die as PISNe, the volume-averaged mean
 metallicity would reach $Z\sim 10^{-4}Z_\odot$ at a redshift of $\sim 12-16$
 \cite{Yoshida04}. It is suggested that the PopIII stars are terminated
 at $z_{\rm end}\sim 9-10$ \cite{Ricotti02b,Salvaterra03,Matsumoto05,CF06}.
 
Based on the above discussions, we assume a PopIII mass fraction, 
 $F_{\rm III}(z)$, which means the mass fraction of 
 objects forming from gas with $Z<Z_{\rm crit}$ at a redshift $z$,
 i.e. the sites of Pop${\hbox{III}}$ star formation\cite{Scannapieco03}.
 Here $F_{\rm III}(z)$ is merely a function of redshift $z$.
 The transition from PopIII to PopII/I should not happen suddenly at a certain
 redshift, because chemical feedback is a local process: with regions close to
 star formation sites rapidly becoming metal-polluted and overshooting $Z_{\rm crit}$,
 and others remaining essentially metal-free. PopIII and PopII star formation
 modes could have been coeval\cite{cf05}, so instead of assigning
 $F_{\rm III}(z)$ a Heaviside function $\theta (z-z_{\rm tran})$, here we introduce
 a function: $F_{\rm III}(\beta,z)=(z/z_{\rm tran})^{\beta}/[1+(z/z_{\rm tran})^{\beta}]$,
 in which $\beta$ is a free parameter describing the transition speed. The $\beta\to \infty$
 limit reproduces the sharp transition case. In this work, we take the transition
 redshift to be $z_{\rm tran}\simeq 14$, and the free parameter $\beta=12$,
 which ensures the mass fraction of PopIII is lower than 5 percent at
 $z_{\rm end}$. Now the SFRD can be rewritten as
\bea
\dot{\rho}_{\rm III}(z)=\int\limits_z^{\infty}\de z_{\rm c}
\int\limits_{M_{\rm low}}^{\infty}\de M'N(M',z,z_{\rm c})\nonumber \\
\times \dot{M}_{\rm SF}(M',z,z_{\rm c})\times F_{\rm III}(z_{\rm c})
\label{sfrd3a}
\eea
for PopIII, and
\bea
\dot{\rho}_{\rm II}(z)=\int\limits_z^{\infty}\de z_{\rm c}
\int\limits_{M_{\rm low}}^{\infty}\de M'N(M',z,z_{\rm c})\nonumber \\
\times \dot{M}_{\rm SF}(M',z,z_{\rm c})\times (1-F_{\rm III}(z_{\rm c}))
\label{sfrd2a}
\eea
for PopII/I.

\section{Mechanical feedback processes}
\subsection{PopII/I SNe feedback}
\begin{table}
\begin{center}
\caption{Model parameters}
\begin{tabular}{r l }
\hline
$Parameters$ & $Referrence \ Values$ \\
 \hline
$\epsilon^{\rm {III}}_{\rm SN}$& 0.05 \\
$\epsilon^{\rm {II}}_{\rm SN}$& 0.02 \\
$f^{\rm {III}}_{\rm esc}$& 0.8\\
$f^{\rm {II}}_{\rm esc}$& 0.1\\
$f^{\rm {III}}_{*}$& 0.07\\
$f^{\rm {II}}_{*}$& 0.3\\
\hline
\end{tabular}
\label {paratable}
\end{center}
\begin{flushleft}
\end{flushleft}
\end{table}
Mechanical feedback is associated with mechanical energy injection from SNe
 explosions and galactic winds. Most of works concentrate on the effects of the
 first-generation SNe explosions at very high redshift rather than the winds
 from metal-free stars \cite{MFM02,Yoshida03,SFS03,KY05,Greif07}.
 A consequence of SNe explosions is to expel the gas out of the
 host halo partially ($blow out$) or totally ($blow away$) and reduce or empty
 the reservoir for subsequent star formation. Some numerical simulations show
 that one PopIII PISN $M_*=200M_{\odot}$ $(E_{\rm SN}\sim 10^{52}\rm{ergs})$
 can deplete its host halo easily(discussed in the next subsection).
 But there are at least two reasons for an SN explosion of PopII/I not being
 capable of blowing the most gas out of its host halo: $(i)$ its explosion
 energy is much lower than that of a PopIII star, and the disperse distribution
 of the lifetime of PopII/I stars cannot ensure enough number of the SNe per
 unit time to deplete the halo; $(ii)$ a halo with virial temperature
 $T_{\rm vir}\ge10^4K$ in the PopII/I epoch($z\le 14$) having mass of
 $M_{\rm vir}\simeq 10^8M_{\odot}$, corresponding to a binding energy
 $E_{\rm b}\simeq10^{52}\sim10^{53}ergs$, is able to prevent an SN explosion
 of PopII/I (generally $E_{\rm SN}\le 10^{51}ergs$) from expelling most gas out
 of the halo, because for halos with mass $M_{\rm vir}\ge 10^7M_{\odot}$ the
 star formation will not be quenched even if $E_{\rm SN}$ exceeds the binding
 energy of halos $E_{\rm b}$ by 2 orders of magnitude\cite{KY05}. Furthermore,
 the enriched metallicity can enhance the gas cooling, which can also help to
 hold the escaping gas back.

However, the huge energy generated by a (PopII/I) SN explosion can, at least,
 partially heat the cold gas even in a high-metallicity
 environment. This normal SNe feedback has been extensively studied both
 in terms of SNe explosions and galactic outflows. Due to the feedback of
 supernovae explosions, the gas will be removed from the cold phase at
 the rate \cite{Granato04}:
\bea
\dot M_{\rm II}^{\rm SN} ~= ~\frac{2}{3} \dot{M}_{\rm SF}(M,z,z_{\rm c}) \,
 \epsilon_{\rm SN}^{\rm II} \, \frac{\eta_{\rm SN}^{\rm II}
 E_{\rm SN}^{\rm II}}{\sigma^2}\nonumber \\
  ~=~ 1.1 \epsilon_{\rm SN}^{\rm II} \dot{M}_{\rm SF}(M,z,z_{\rm c})
 \, \left(\frac{500 \hbox{km}\,
\hbox{s}^{-1}}{V_{\rm c}}\right)^2,
 \label{snfb}
\eea
where $\eta_{\rm SN}^{\rm II}\simeq6\times10^{-3}M_{\odot}^{-1}$ is the number
 of SNe expected per solar mass of formed stars,
 $E_{\rm SN}^{\rm II}$ is the kinetic energy of the ejecta from each PopII/I
 supernova ($6\times10^{50}\,$erg; e.g. Nadyozhin 2003), and
 $\epsilon_{\rm SN}^{\rm II}$ is the fraction of this energy that is used
 to heat the cold gas. Here $\eta_{\rm SN}^{\rm II}$ is evaluated by 
 ``Starburst99" and adopting a minimum progenitor mass of $8\,M_{\odot}$ 
 and the Salpeter IMF.
 Some analyses show that above 90\% of the
 SN kinetic energy may be lost by radiative cooling \cite{Thornton98,Heckman00}.
 Here we set $\epsilon_{\rm SN}^{\rm II}=0.02$.
 As the line-of-sight velocity dispersion $\sigma$,
 we adopt the relationship $\sigma \simeq 0.65 V_{\rm c}$
 \cite{Ferrarese02}.
Finally, we can obtain the SNe feedback on PopII/I SFRD:
\bea
\dot{\rho}_{\rm II}^{\rm FB}(z)  =  \int\limits_z^{\infty} \de z_{\rm c}
\int\limits_{M_{\rm low}}^{\infty} \, \de M'\ N(M',z,z_{\rm c})\nonumber \\
 \times \bigg[1-F_{\rm III}(z_{\rm c})\bigg]
 \times\dot{M}_{\rm SF}^{\rm II}(M',z,z_{\rm c}) .
\label{sfrd2HI}
\eea
where  $\dot{M}_{\rm SF}^{\rm II}(M,z,z_{\rm c})$ is the regulated SFR by the
 feedback from SNe explosions in PopII/I dominated halos (with $T_{\rm vir}\ge 10^4 K$,
 see $Appendix$ A for detail).


\subsection{ Mechanical feedback from first-generation stars}
\label{PopIII SNFB}
 The ultimate fate of a metal-free star depends critically on
 its mass \cite{heger02,heger03}: $(1)$ $8M_{\odot}< M_*< 25M_{\odot}$ (these
 stars explode as core-collapse SNe and leave neutron stars behind),
 $(2)$ $25M_{\odot}< M_*< 40M_{\odot}$ (these explode as faint Type II SNe and
 leave black holes behind), $(3)$ $40M_{\odot}< M_*< 140M_{\odot}$ (these do not
 explode as SNe and directly collapse into black holes\footnote{ But some
 of them experience a pulsating instability and eject their outer envelope,
 again leaving
 black holes behind.}), $(4)$ $140M_{\odot}< M_*< 260M_{\odot}$ (these explode as
 PISNe, causing complete disruption),
 $(5)$ $260M_{\odot}< M_*$ (these collapse, in the absence of rotation, directly
 into black holes), so not all the PopIII stars can die as an SN explosion.
 We employ a slightly top-heavy IMF (Larson, 1998) for PopIII stars
($50M_{\odot}\le M_* \le 500M_{\odot}$):
\bea
 \f{\de N}{(\de{\rm log}M_*)} \propto (1+M_*/M_{\rm c})^{-1.35},
\label{IMF3}
\eea
 where $M_{\rm c}=100M_{\odot}$ is the characteristic stellar mass of PopIII.
 As a result, the PISN is almost the only type of supernova explosion for PopIII.
 Bromm, Yoshida \& Hernquist (2003)
 show that, for a halo of mass $M\sim 10^6M_{\odot}$ at $z\sim 20$, a PISN of
 mass $M_*=250M_{\odot}$$(E_{\rm SN}\sim 10^{53}\rm{ergs})$
 can disrupt the halo completely. A similar result has been obtained
 by Greif et al. (2007), who find that a PISN with mass
 $M_*=200M_{\odot}$$(E_{\rm SN}\sim 10^{52}\rm{ergs})$ can
 deplete the whole host halo.

The photoevaporation effect might be particularly important for PopIII objects
 \cite{cf05}. In small-halo objects($T_{\rm vir}< 10^4 K$), photoevaporation
 alone from OB stars can produce strong galactic winds that expel most of
 the gas from galaxies. Galactic winds produced by an SN explosion may
 be important after about 10 Myr\cite{Ricotti08}. Because the internal
 photoevaporation in a PopIII object is able to deplete most of the gas before
 the first SNe explode, in fact, the first SNe explosions have no chance to
 exert negative feedback on the star formation in these halos.
 Finally we only consider the SNe feedback in halos
 of $T_{\rm vir}> 10^4 K$ in this paper. For those big halos($T_{\rm vir}> 10^4 K$),
 the high cooling effect and deeply gravitational well can confine the gas
 photoevaporation. And even the most powerful PISN ($E_{\rm SN}\simeq10^{53}ergs$)
 cannot significantly blow the gas away\cite{KY05}.
 In reference to the mechanism of the SNe feedback of PopII/I,
 we use the same formula but with different parameters to describe
 the PopIII-SNe feedback in these halos ($M> 10^8M_{\odot}$),
 \bea
\dot M_{\rm III}^{\rm SN} = \frac{2}{3} \dot{M}_{\rm SF}(M,z,z_{\rm c}) \,
 \epsilon_{\rm SN}^{\rm III} \,
 \frac{\eta_{\rm SN}^{\rm III} E_{\rm SN}^{\rm III}}{\sigma^2}\nonumber \\
  = 4.5 \epsilon_{\rm SN}^{\rm III} \dot{M}_{\rm SF}(M,z,z_{\rm c}) \,
 \left(\frac{500 \hbox{km}\,
\hbox{s}^{-1}}{V_{\rm c}}\right)^2.
 \label{snfb3}
\eea
Here $E_{\rm SN}^{\rm III}\simeq10^{52}ergs$,
 $\eta_{\rm SN}^{\rm III}\simeq1.5\times10^{-3}M_{\odot}^{-1}$(evaluated with
 the PopIII IMF and mass range). For lack of information about the strength
 factor of the SNe feedback of PopIII, we set
 $\epsilon_{\rm SN}^{\rm III}=0.05$\cite{Granato04,Lapi06} as our reference value.
 One can obtain the PopIII SFRD with SNe feedback:
\bea
\dot{\rho}^{\rm FB}_{\rm III}(z)=\int\limits_z^{\infty}\de z_{\rm c}
F_{\rm III}(z_{\rm c})\int\limits_{M(z_{\rm c},10^4K)}^{\infty}
N(M',z,z_{\rm c})\nonumber \\ 
\times\dot{M}_{\rm SF}^{\rm III}(M',z,z_{\rm c})\de M'
\label{sfrd3HI}
\eea
{where $\dot{M}_{\rm SF}^{\rm III}(M,z,z_{\rm c})$ is the regulated SFR by the
 feedback from SNe explosions in PopIII halos of 
 $T_{\rm vir}>10^4K$(see $Appendix$ A for detail). Of course, it is definite that
 the number density of halos with $T_{\rm vir}>10^4K$ is tiny in PopIII epoch
 according to the hierarchical clustering scenario.}

\begin{figure*}
\centerline{
\psfig{figure=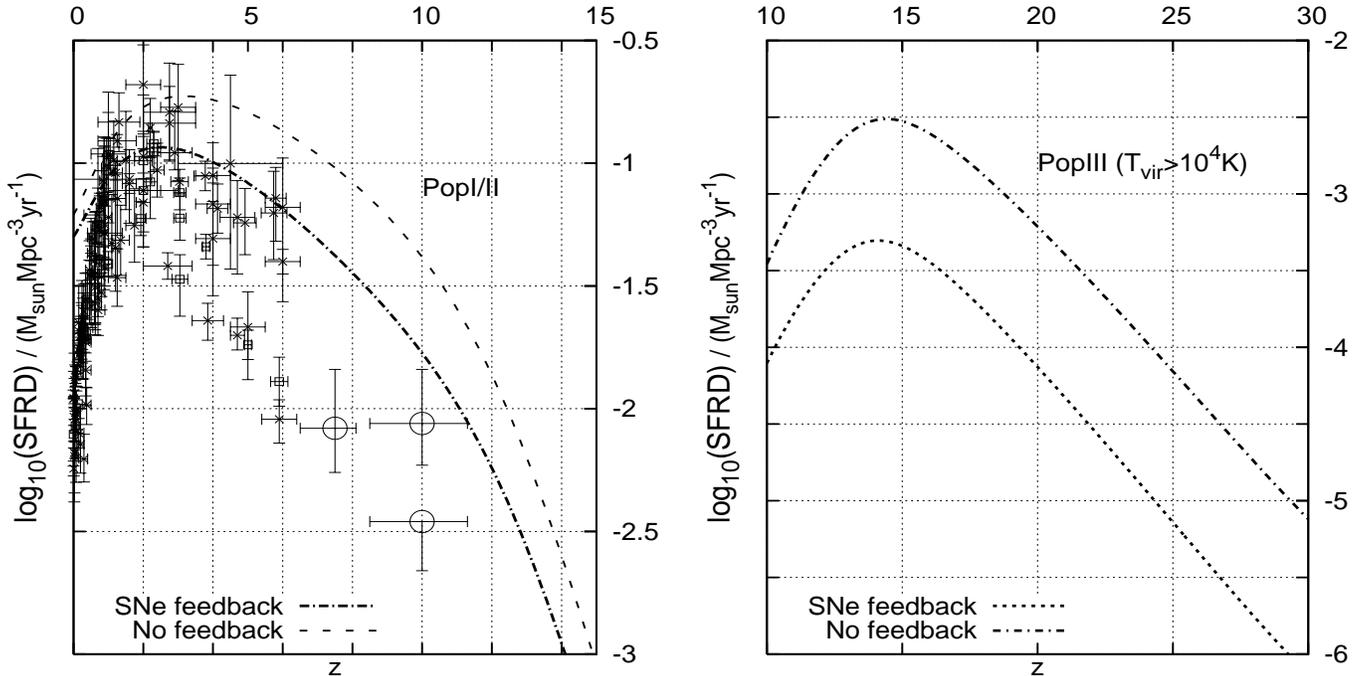,height=9.5cm,width=\textwidth,angle=270.0}}
\caption[]{Evolution of SFRD with the redshift $z$. 
{\it Left panel:} the crossed data points are taken from
 Hopkins \& Beacom (2006). The squared ones come from Reddy et al. (2008). 
 Three circled points are from the UDF data (Bouwens et al. 2005).
 The dashed line represents the PopII/I SFRD without SNe feedback, while the
 dot-dashed one with the SNe feedback. {\it Right panel:} the original
 SFRD of PopIII ($T_{\rm vir}>10^4K$, the dot-dashed line) is suppressed down about
 one order of magnitude by the SNe feedback (the dotted line). }
\label{SNFB}
\end{figure*}

$\hbox{{Figure\ \ref{SNFB}}}$ shows the SNe feedback in PopIII($T_{\rm vir}>10^4K$)
 and PopII/I halos. In the left panel, we scale down all the observation
 data by a factor 1.9 to make it consistent with the IMF used in our model
 ($M^{\rm min}_*$=0.5$M_\odot$, $M^{\rm max}_*$=100$M_\odot$
 for Pop${\hbox{II/I}}$ stars and Salpeter type IMF).
 In the right panel, the original SFRD of PopIII($T_{\rm vir}>10^4K$)
 is cut down about one order of magnitude by the SNe feedback.  

 $\hbox{{Figure\ \ref{SNFB}}}$ manifests the relative feedback strength of SNe
 feedback[$\dot{\rho}^{\rm FB}_{\rm III}(z)/\dot{\rho}_{\rm III}(z)$ or
 $\dot{\rho}^{\rm FB}_{\rm II/I}(z)/\dot{\rho}_{\rm II/I}(z)$]
 reduces as the redshift decreases, because massive halos appear
 abundant only at low redshift according to the ``bottom-up" hierarchical 
 structure
 model (see Mo \& White 2002 for details) and massive halos whose binding energy
 is much higher than before can reduce the feedback effect from SNe explosions.
 On the other hand, massive stars disappear as the metallicity
 of gas is enriched. Small star formation 
 with a long lifetime and small $E_{\rm SN}$ is enhanced in the
 metal-enriched cloud.

\section{Radiative feedback and cosmic reionization}

The ionizing radiation produced by massive stars or quasars can have
 local effects or long-range effects, either affecting the formation and evolution
 of nearby objects or joining the radiation produced by other galaxies to form a
 background. Mini-QSOs that are usually thought to be powered by accretion onto
 the first black holes may provide an X-ray background when the universe has a
 certain number of mini-QSOs in a unit comoving volume. This background would
 ionize the HI to $H^+$ and $e^-$ by direct photoionization or collisional
 excitation from a high-energy photoelectron produced by photoionization. If this
 event occurs at the core of a protogalaxy, the fraction of molecular hydrogen
 would be promoted\cite{Haiman00,Glover03} via
\begin{eqnarray}
 \rm H + \rm e^- & \rightarrow & \rm H^{-} + \gamma,  \\
 \rm H^{-} + \rm H & \rightarrow & {\rm H_2} + {\rm e^-}.
 \label{h2f1}
\end{eqnarray}
 Dijkstra et al.(2004) and Salvaterra et al.(2005) show that
 the hard X-ray from the same sources will produce a present-day soft X-ray
 background. They find that the models with accreting black holes will overproduce
 the observed X-ray background by a large factor. A population dominated by
 mini-QSOs could still partially ionize the IGM at $z>6$,but its contribution
 can be severely constrained if the X-ray background is resolved further into
 discrete sources. By considering the constraint from the
 soft X-ray background, Ricotti \& Ostriker (2004b) show that mini-quasars at
 high-z do not overproduce the X-ray background and can still produce a significant
 contribution to reionization and the optical depth of electrons
 $\tau_{\rm e}$.

 Based on the investigation of the environment and reionization process around
 the highest redshift QSOs having Gunn-Peterson troughs ($z > 6.1$),
 Yu \& Lu(2005) argue that a significant fraction of
 hydrogen in the Stromgren sphere around QSOs is ionized by photons from stars
 and that only about several percent to at most $10\%-20\%$ of the total hydrogen
 is left (e.g., in minihalos, halos, or high-density subregions) to be ionized
 by QSO photons. Willott et al.(2005) show that the current constraints on the
 quasar population give an ionizing photon density $\ll 30\%$
 that of the star-forming
 galaxy population by analyzing the observational data from
 the Canada-France-Hawaii Telescope. They conclude that active galactic nuclei
 make a negligible contribution to the reionization of hydrogen at $z\sim 6$.
 But this argument only applies to quasars at $z\le 6$, one can still
 have a high$-z$
 population of mini-quasars that partially ionize the IGM at $z>10$ without
 overproducing the X-ray background\cite{Ricotti05}.

Pre-ionization by X-rays can increase the IGM optical depth from
 $\tau_{\rm e}\simeq 0.06$ given by stellar sources only to
 $0.1\le \tau_{\rm e}\le 0.2$\cite{Ricotti04b}. From $4\sim 5$ years ago,
 it was necessary to fit an observational value of
 $\tau_{\rm e}\simeq 0.17$ measured by WMAP satellite.
 Recent WMAP data indicating $0.068\le \tau_{\rm e}\le 0.1$ maybe imply that
 pre-ionization by X-rays is not as important as before. In this paper, we focus
 on the ionization by stellar sources and ignore the effects from QSOs or
 mini-QSOs. We will consider the AGN feedback carefully in the work in
 preparation.

\subsection{Photoevaporation in small-halo objects ($T_{\rm vir}<10^4K$)}

Pop${\hbox{III}}$ stars quite likely reside in small DM halos and a
 metal-free environment. It benefits the photons escaping
 and the radiative cooling. As mentioned in the last section, photoevaporation
 is important in small-halo objects
 ($T_{\rm vir}<10^4K$, e.g. most of the Pop${\hbox{III}}$ halos). It can prevent
 the size of HII regions from exceeding
 $R^{\rm com}_{\rm HII} \simeq 5h^{-1}{\rm kpc}$, the mean free path of ionizing
 photons, about the size of the dense filaments and the virial radii of the
 halos \cite{Ricotti02b}. When the HII regions become bigger than the
 filaments, molecular hydrogen is destroyed and
 the star formation is suppressed. An analytic
 model has been developed to describe the propagation of ionization 
 fronts in the IGM \cite{bl01}:
\bea 
<n^{\rm H}_{\rm ISM}> \frac{\de V_{\rm p}}{\de t}= \frac{\de N_\gamma}{\de t} 
- \alpha_{\rm B} \left<(n^{\rm H}_{\rm ISM})^2\right> V_{\rm p}, 
\label{front} 
\eea
where $N_\gamma$ is the number of ionizing photons produced by the source.
Here $V_{\rm p}$ is the ionized proper volume,
$\alpha_{\rm B}\simeq 2.6\times 10^{-13}$cm$^3$ s$^{-1}$ is the case B
 recombination coefficient at $T \simeq 10^4$ K \cite{Seager99,bl01},
 and $n^{\rm H}_{\rm ISM}$ is the proper number density of the hydrogen atoms.
 We assume the ISM in halos is homogeneous, which means
 $<(n^{\rm H}_{\rm ISM})^2>=<(n^{\rm H}_{\rm ISM})>^2$,
 so the analytical solution is
 \bea
V_{\rm p}(t,t(z_c))=\int_{t(z_c)}^t \frac{1}{<n^{\rm H}_{\rm ISM}(t')>}
 \frac{\de N_{\gamma}}{\de t'} \,\nonumber \\
\times \exp\left\{{-\alpha_{\rm B}
\int_{t'}^t {<n^{\rm H}_{\rm ISM}(t'')>}\, \de t''}\right\}
\de t'\ ,
\label{HIIsoln} 
\eea 
where\footnote{The mass fraction of baryons in a halo is
 $\f{\Omega_{\rm b}}{\Omega_{\rm m}}$, so the baryon density in halos is
 at least $\Delta_c(t)\times \frac{\Omega_{\rm b}}{\Omega_{\rm m}}$ times the
 one in IGM.}
\bea 
<n^{\rm H}_{\rm ISM}(t)>=\Delta_c(t)\times \f{\Omega_{\rm b}}{\Omega_{\rm m}}
\times \bar{n}^0_{\rm H}\times(1+z(t))^3, \\
\bar{n}^0_{\rm H}=1.88\times10^{-7} \left(\frac{\Omega_b h^2}{0.022}\right)\
 {\rm cm}^{-3},\nonumber
\eea
and $\bar{n}^0_{\rm H}$ is the present number density of hydrogen.
 As mentioned in Eq.~(\ref{tdyn}), $\Delta_c$ is the overdensity relative to the
 critical density at the collapse redshift.
 Following Barkana \& Loeb (2001), we can evaluate ${\de N_{\gamma}}/{\de t}$ by
using
\bea 
\frac{\de N_{\gamma}}{\de t}=\frac{\alpha-1}{\alpha} \frac{n^{\rm III}_{\gamma}}{t_s},
\label{dngdt}
\eea
 where $\alpha$ is the index in the mass-luminosity
 relation (for OB stars $\alpha=4.5$), and $n^{\rm III}_{\gamma}$
 the number of ionizing photons released per baryon of stars formed
\cite{schaerer03,Haiman06}. {Fang $\&$ Cen (2004) present a relationship
 between $n^{\rm III}_{\gamma}$ and the stellar mass, $M_*$.
 Because $n^{\rm III}_{\gamma}$ is not sensitive to $M_*$ in the mass
 range of concern, we set $n^{\rm III}_{\gamma}$ as a fixed value
 $80000$ in this paper\cite{FC04}.}
 Most massive stars fade away with the
 characteristic time scale $t_s=3\times 10^6$yr\cite{bac1984,abs06}.
For a halo of mass $M$, star formation initiated at $t(z_c)$, the proper
 size(or radii) of the HII region at later time $t$, is 
\begin{figure}
\centerline{
\psfig{figure=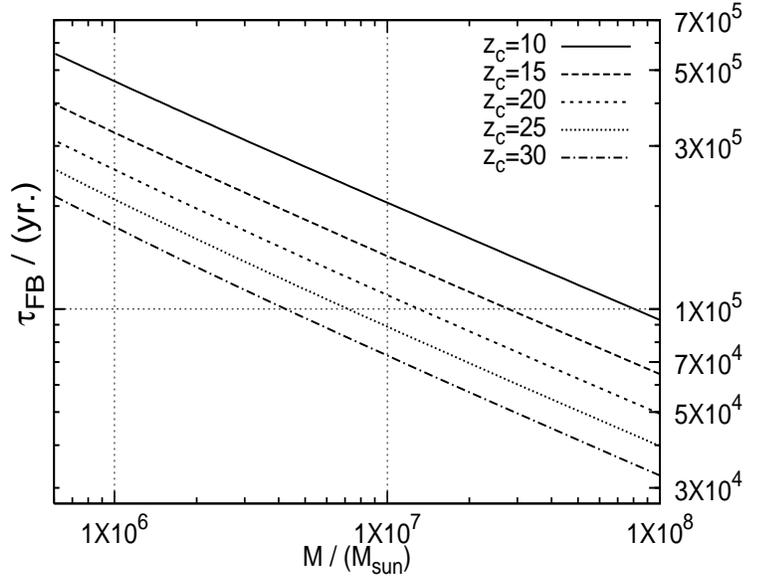,height=8.0cm,width=10cm,angle=270.0}}
\caption[]{The duration time of star formation $\tau_{\rm FB}=t'-t'(z_c)$
 versus the masses of halos.}
\label{tauM}
\end{figure}
\begin{equation}
R(M,t(z_c),t)=\left\{\frac{3}{4\pi}f_*^{\rm III} f_{\rm esc}^{\rm III}
 \frac{\Omega_{\rm b}M}
{\Omega_{\rm m}m_{\rm p}}\times V_{\rm p}(t,t(z_c))\right\}^{1/3},
\end{equation}
where $f_{\rm esc}^{\rm III}$ is the escape fraction of ionizing photons
 from the sources(or resolution elements in 3D numerical simulations,
 see Ricotti et al. 2002a, 2002b, 2008), and $f^{\rm III}_{\rm esc}\simeq 0.3\sim 0.9$
 is suggested by Choudhury \& Ferrara (2006) and Mao et al.(2007). Yoshida et
 al.(2007) provide a time-averaged ionizing photon escape fraction as a function
 of stellar mass $M_*$. For the mass range adopted in this work, we use a fixed
 value $f^{\rm {III}}_{\rm esc}=0.8$.
 As mentioned above, star formation will be suppressed when $R(M,t'(z_c),t')$ is
 equal to the size of the filaments. Thus the relation between $t'-t'(z_c)$ and $M$
 is given by
\bea
R(M,t'(z_c),t')=R^{\rm com}_{\rm HII}/(1+z').
\eea
We illustrate this relationship in $\hbox{{Fig.\ \ref{tauM}}}$,
 where $\tau_{\rm FB}=t'-t'(z_c)$ is the duration time of star formation.
 We plot five cases $z_c=10,15,20,25,30$, which
 are actually not straight lines in the figure. 
 For the halos with the same mass, $\tau_{\rm FB}$
 increases as the redshift decreases. The bigger the halo, the less time it
 spends enlarging the HII region to the size of the filaments.
 If $\tau_{\rm FB}<t_s$, the photoevaporation feedback occurs
 before the first SN exploded. And the inequality
 $c\times \tau_{\rm FB}>R^{\rm com}_{\rm HII}/(1+z')$ ($c$ the speed of light)
 is valid for all these cases, which means the speed of the propagation of the
 ionization fronts cannot exceed the speed of the light\footnote{However,
 it seems that the offcenter sources may make a superluminal spread
 when their HII regions get an overlap. But it is not real superluminal
 behavior.}.

\begin{figure}
\centerline{
\psfig{figure=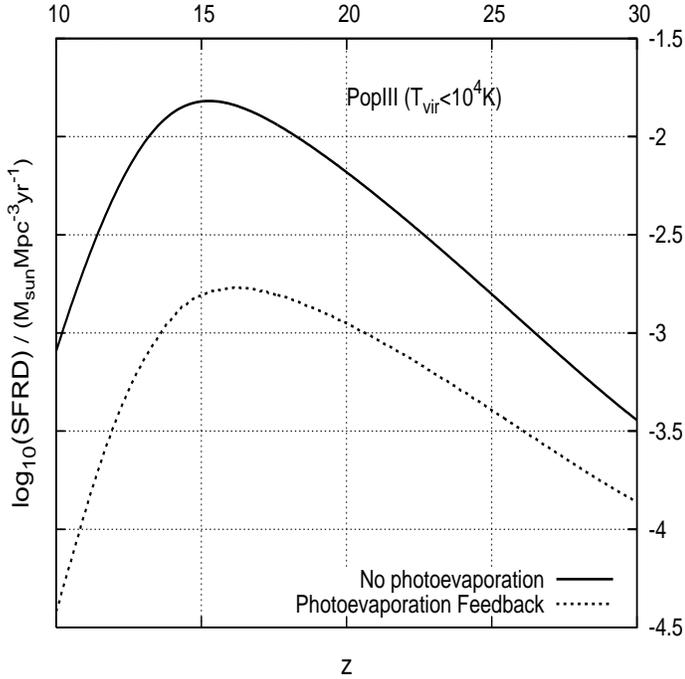,height=9.5cm,width=9.5cm,angle=270.0}}
\caption[]{Evolution of SFRD as a function of the redshift $z$ with/without the
 photoevaporation feedback effect. The solid line represents the SFRD
 without feedback. The dotted one stands for the case that the SF is suppressed
 by photoevaporation. Both of them represent the PopIII SFRD in halos with
 $T_{\rm vir}<10^4 K$.}
\label{PEFB}
\end{figure}

Now we can calculate the SFRD in PopIII halos with $T_{\rm vir}<10^4 K$ by using

\bea
\dot{\rho}^{\rm FB}_{\rm III}(z,<10^4 K)=\int\limits_z^{\infty}\de z_{\rm c}
F_{\rm III}(z_{\rm c})\int\limits_{M(z_c,10^3 K)}^{M(z_c,10^4 K)}
N(M',z,z_{\rm c}) \nonumber \\
\times \theta(t(z_{\rm c})+\tau_{\rm FB}-t_{\rm z})\times  \dot{M}_{\rm SF}
(M',z,z_{\rm c})\de M' 
\label{sfrd3l4}
\eea
where $\theta$ is a Heaviside function. $\hbox{{Figure\ \ref{PEFB}}}$ shows the effects
 of this kind of feedback:$(i)$ they suppress the PopIII SFRD down about one order of
 magnitude at the peak ($z\sim 15$);$(ii)$ photoevaporation gets stronger when
 redshift decreases. And we also find that the feedback will become weak
 when $f_{\rm esc}^{\rm III}$ is tuned down. Moreover, the SFRD is not
 sensitive to the $f_*^{\rm III}$.

\subsection{Radiative feedback in large halos ($T_{\rm vir}>10^4 K$)}

For halos with $T_{\rm vir}$ higher than the cooling temperature of the hydrogen
 atoms($10^4 K$), gas cools via HI emission lines, and the photoevaporation by
 internal sources cannot expel gas significantly. But the ionizing background can
 still suppress some small halos in which the star formation is about to initiate.
 Due to the IGM reionized by UV photon escape from star-forming
 galaxies, the temperature of the gas in halos is enhanced.
 This can dramatically increase the Jeans mass.
 Furthermore, numerical simulations indicate that the photoionizing background
 can completely suppress galaxy formation in halos with circular velocity
 $V_{\rm c}\le 35$~km~s$^{-1}$,
 while the mass of cooled baryons is reduced by 50\% for  halos with circular
 velocities $\sim 50$~km~s$^{-1}$ \cite{tw96}.
 In the ionized fraction of the universe, we assume complete
 suppression of star formation in halos below circular velocity
 $V_{\rm c}=35$~km~s$^{-1}$ and no suppression above circular velocity of
 $95~$km s$^{-1}$. For intermediate masses, we adopt a linear fit from $1$
 to $0$ for the suppression factor (as in Bromm \& Loeb, 2002).
  Thus in the HII region the SFRD can be expressed by
\bea
\dot{\rho}_{\rm II}^{\rm HII}(z)  =  \int\limits_z^{\infty} \de z_{\rm c}
\int\limits_{M(z_c,10^4 K)}^{\infty} \, \de M'\ N(M',z,z_{\rm c}) \nonumber \\
 \times \bigg[1-F_{\rm III}(z_{\rm c})\bigg]
 \times\dot{M}_{\rm SF}^{\rm II}(M',z,z_{\rm c})
\times W(V_{\rm c}) .
\label{sfrd2HII}
\eea
 for PopII/I and
\bea
\dot{\rho}^{\rm HII}_{\rm III}(z,>10^4 K)=\int\limits_z^{\infty}\de z_{\rm c}
F_{\rm III}(z_{\rm c}) \int\limits_{M(z_c,10^4 K)}^{\infty}
N(M',z,z_{\rm c})\nonumber \\
\times \dot{M}_{\rm SF}^{\rm III}(M',z,z_{\rm c})\times W(V_{\rm c})\de M'
\label{sfrd3HII}
\eea
for PopIII, where $W(V_{\rm c})$ is 
\bea
W(V_{\rm c})= \begin{cases} 
0 ,&\hbox{$V_{\rm c}\le 35\hbox{km}\,\hbox{s}^{-1}$}\\
(V_{\rm c}-35)/(95-35),&\hbox{$35
\le V_{\rm c}\le 95\hbox{km}\,\hbox{s}^{-1}$}\\
1 ,&\hbox{$V_{\rm c}\ge 95\hbox{km}\,\hbox{s}^{-1}$}
\end{cases}.
\label{WVC}
\eea

\begin{figure*}
\centerline{
\psfig{figure=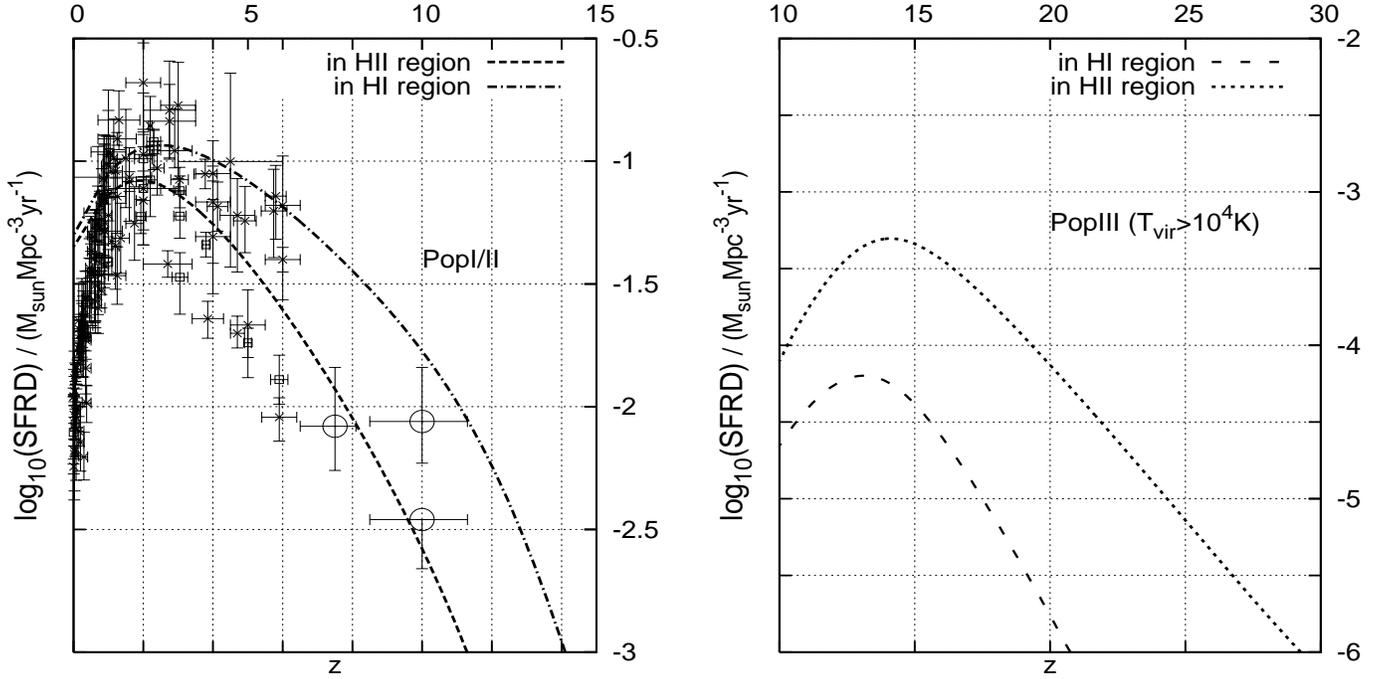,height=9.5cm,width=\textwidth,angle=270.0}}
\caption[]{Evolution of SFRD in two extreme cases: the stars form in
 the completely reionized universe (HII region, the dashed line) and in the
 completely neutral universe (HI region, the dot-dashed line in the lefthand
 panel and the dotted line in the right one). Observational data are the
 same as Fig.~\ref{SNFB}.}
\label{HIHII}
\end{figure*}

We plot two extreme cases in $\hbox{{Fig.\ \ref{HIHII}}}$: the stars form in the
 completely reionized universe in HII region and in the
 completely neutral universe in HI region. 
 This radiative feedback from
 PopII/I($left\ panel$) and PopIII has the same behavior: feedback gets stronger
 when redshift increases. {Similar to the SNe mechanical feedback mentioned above,
 the more massive the halo, the less the SF can be affected.}

 For the ionized
 fraction of the universe at a given redshift $z$, it can be described by the
 following equation. For simplicity, we assume that all the Lymann continuum
 photons escape from a star-forming galaxy are involved in reionizing
 the IGM, and the photons emitted from the sources will immediately join the
 action with the atom, regardless of the photon propagation between the source
 and the atom\footnote{Recent work shows that the finite speed of light will
 lead to a substantial change of the growth rate in the ionized volume
 (see Qiu et al. 2008 for more details).}. Then the fraction of ionized
 hydrogen $Q_{\rm HII}$, the so-called filling factor, evolves as \cite{bl01},
\begin{equation}
\f {\de Q_{\rm HII}}{\de z } = \f{f_*f_{\rm esc}}{(1+z)^3\bar{n}^0_{\rm H}}
\f {\de{N}_{\gamma}}{\de t}~\f{\de t}{\de z} -\alpha_{\rm B}(1+z)^3
\bar{n}^0_{\rm H}Q_{\rm HII} C~\f{\de t }{\de z},
\label{diffeq}
\end{equation}
where the volume-averaged clumping factor of the IGM, $C$, is defined as
 $C\equiv \langle n^2_{\rm H} \rangle / ((1+z)^3{\bar n}_{\rm H}^0)^2$.
 The first term on the right is the rate of ionization
 and the second term is the rate of recombination, weighted by the $Q_{\rm HII}$,
 as recombinations take place only in the ionized region.
 Then ${\de{N}_{\gamma}}/{\de t}$ is obtained from the SFRD calculation as,

\bea
\f {\de{N}_{\gamma}}{\de t}= \f{\dot{\rho}
_{\rm SF}(z,>10^4 K)(1+z)^3}{f_*m_{\rm p}} n_\gamma ,
\label{dotng}
\eea
where $\dot{\rho}_{\rm SF}(z,T_{\rm vir}>10^4 K)$ is the SFRD in
 halos of $T_{\rm vir}>10^4 K$ (including all PopII/I halos and a few
 PopIII halos) at $z$. Because the strong photoevaporation feedback in small-halo
 objects can quench the SF, which stops the ionizing photons to escape to the IGM.
 This part of SFRD contribute little to the IGM reionization.
 The value of $n_\gamma$ depends on the IMF of the forming stars. For a Salpeter
 IMF(with $0.5~M_{\odot} \le M_* \le 100~M_{\odot}$), $n^{\rm II}_\gamma$
 is about $5,400$, evaluated by ``Starburst99"
\footnote{http://www.stsci.edu/science/starburst99/} with metallicity $Z=0.008$,
 however, for the metal-free stars, the IMF could
 be biased towards very massive stars. As mentioned in the last section,
 we still used  $n^{\rm III}_\gamma \sim 80,000$ as our reference
 value. For the clumping factor of IGM, $C$, we adopted the simple form given by
 Haiman \& Bryan (2006).

\begin{figure*}
\centerline{
\psfig{figure=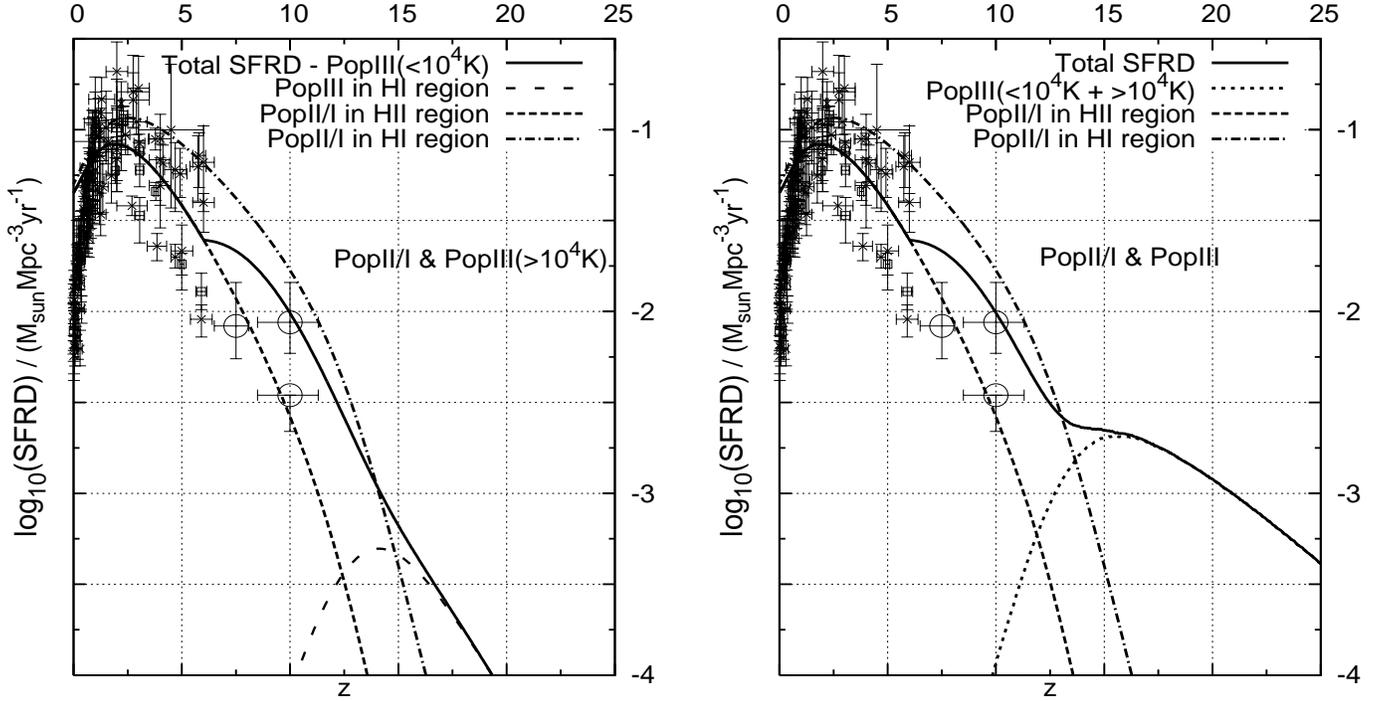,height=9.5cm,width=\textwidth,angle=270.0}}
\caption[]{Evolution of SFRD with feedback. {\it Right panel:} the solid line
 stands for the SFRD in halos of $T_{\rm vir}>10^4 K$. Other lines are the same
 ones as in $\hbox{{Fig.\ \ref{HIHII}}}$. 
 {\it Left panel:} The PopIII SFRD in halos of $T_{\rm vir}<10^4 K$.}
\label{RFB}
\end{figure*}

\begin{figure*}
\centerline{
\psfig{figure=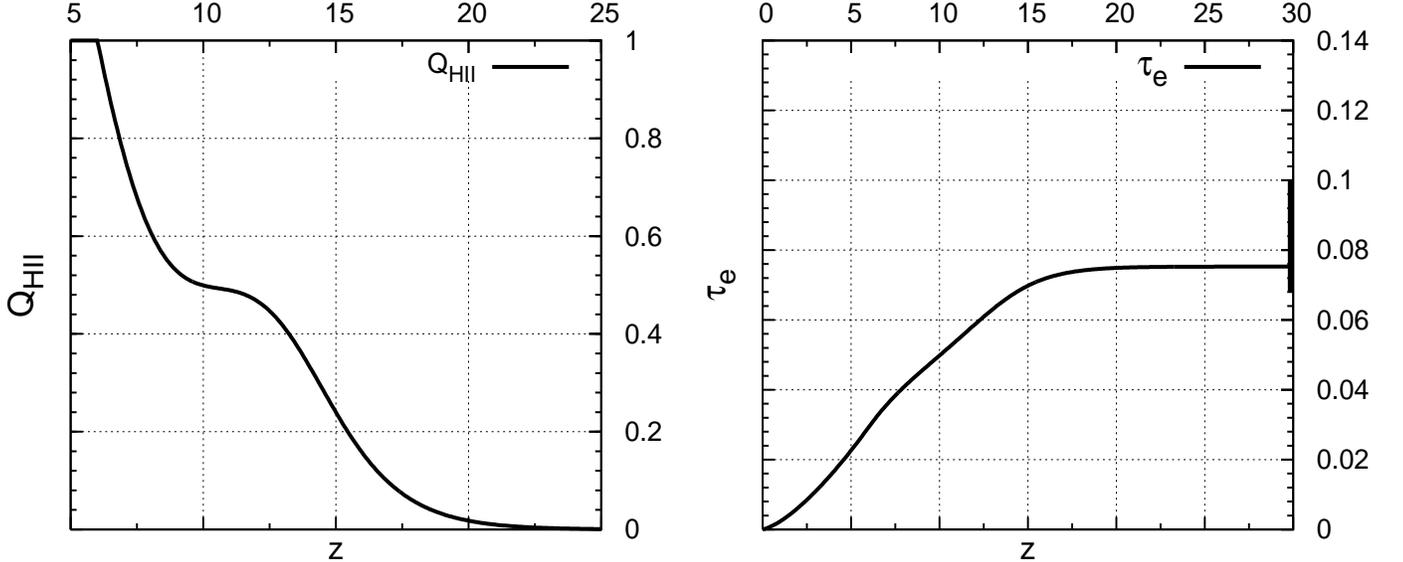,height=8cm,width=\textwidth,angle=270.0}}
\caption[]{Evolution of the filling factor
 $Q_{\rm HII}$ ({\it left panel}) and the optical
 depth to Thomson electron scattering $\tau_e$ ({\it right panel}) with the redshift. }
\label{reiontaue}
\end{figure*}

By solving Eq.~(\ref{diffeq}) (see $Appendix$ B for detail), one can find
 that SF and reionization affect each other.{When star formation gets high,
 HI ionization becomes strong and star formation is suppressed. Hopefully, $JWST$
 can help us to understand their relationship (SF and reionization) exactly.
 The result is plotted in $\hbox{{Fig.\ \ref{RFB}}} (left\ panel)$.
 One may find that the solid line separates from the
 SFRD(PopII/I+PopIII) in HI region and eventually merges into the SFRD line
 in HII region as the filling factor $Q_{\rm HII}$ (see $\hbox{{Fig.\ \ref{reiontaue}}}$)
 increases from zero to unit. The PopIII SFRD in halos of $T_{\rm vir}<10^4 K$
 is taken into account in the $righthand\ panel$
 in $\hbox{{Fig.\ \ref{RFB}}}$. By comparing
 these two panel, we can find that the main part of PopIII SFRD is still contributed
 by those stars in small-halo objects.} Since the PopIII PISNe contaminate the
 IGM with heavy elements at very high
 efficiency (recall the ultimate fate of metal-free stars), PopIII ends its era
 by itself quickly (survival time $\sim 2\times 10^8 yr.$). The ignition of
 PopII stars makes the reionization continue and eventually completes
 it at $z\sim 6$. 

The reionization history and the optical depth of electrons are
 plotted in $\hbox{{Fig.\ \ref{reiontaue}}}$. Here we take the singly
 ionized HeII into account in $righthand\ panel$ in
 $\hbox{{Fig.\ \ref{reiontaue}}}$. The equation for calculating
 the optical depth of electrons is:
\bea
{\tau}_{\rm e}(z) = \int n_{\rm e}(z){\sigma}_{\rm e}\frac{\de s}{\de z}{\de z}
\eea
where $n_{\rm e}(z)$ is the mean electron number density in the universe,
 ${\sigma}_{\rm e}$ stands for the Thompson cross section for electron
 scattering, and $\de s = -c\de z/H(z)(1+z)$ is the proper
 line element. {$\hbox{{Figure\ \ref{reiontaue}}}$ illustrates that the
 model output can fit the observation from the highest QSOs
 ($z_{\rm re}\sim 6$), and WMAP 5 years result
 ($\tau_{\rm e} = 0.084^{+0.016}_{-0.016}$ Komatsu et al. 2008)
 at the same time. Most of the UV photons emitted from small halos that have
 the overwhelming majority in PopIII objects are confined in
 $R^{\rm com}_{\rm HII}$. As shown in the $lefthand\ panel$ in
 $\hbox{{Fig.\ \ref{reiontaue}}}$, PopIII stars are not able
 to reionize the universe significantly (about $20\%$ at $z\sim 14$). In the
 $righthand\ panel$, the optical depth to Thomson scattering provided by 
 PopIII stars is less than $22\%$.}

\section{Discussion and conclusions}
 In this paper we develop an analytic model of the mechanical
 and radiative feedback of star formation based on some results of numerical
 simulations. Basically, our results can fit the
 observational data. However, this model also depends on
 some initial parameters (most of them are listed in $\hbox{Table\ 1}$) and
 working assumptions. 

$(i)$ It is possible to fit the reionization result from the highest QSOs observation
 by providing another pair of $f^{\rm {II}}_{*}$ and $f^{\rm {II}}_{\rm esc}$.
 This is because the IGM reionization mainly depends on
 $n^{\rm II}_\gamma$ and $f^{\rm {II}}_{*}\times f^{\rm {II}}_{\rm esc}$. 
 The value of $n^{\rm II}_\gamma$ is nearly fixed under certain
 conditions (e.g. metallicity, IMF, and stellar mass range),
 but lacking the SFRD observational data constraint
 on $f^{\rm {II}}_{*}$, one can find many pairs of $f^{\rm {II}}_{*}$ and
 $f^{\rm {II}}_{\rm esc}$ by keeping
 $f^{\rm {II}}_{*}\times f^{\rm {II}}_{\rm esc}$
 invariant to guarantee $\tau_{\rm e} = 0.084^{+0.016}_{-0.016}$. 

$(ii)$ As for the assumption on $F_{\rm III}(z)$, there are a lot of similar
 functions. This is an uncertainty factor in this model.
 Hopefully, it may be determined by
 the numerical simulations and even the future JWST observation.

$(iii)$ We only consider the negative feedback in this work,
 the neglect of some positive feedback, such as the X-ray background or dense
 shell (from SNe explosions) fragmentation, results in the underestimation of
 the SFRD, $Q_{\rm HII}$, $\tau_e$, and so on. This is what we will investigate
 in the next work.

$(iv)$ Although we can fit the optical depth to Thomson scattering in this
 work, it does not mean the X-ray pre-ionization is not important. Obviously,
 our results are close to the low fringe of the error bar. Perhaps, the left
 space in the error bar is waiting for the contribution from X-ray
 pre-ionization.
In the following, we summarize the results of this work in more detail.

1. The SNe feedback can suppress the star formation in large halos
 ($T_{\rm vir}>10^4 K$). And the relative feedback strength of SNe feedback
 [$\dot{\rho}^{\rm FB}_{\rm III}(z)/\dot{\rho}_{\rm III}(z)$ or
 $\dot{\rho}^{\rm FB}_{\rm II/I}(z)/\dot{\rho}_{\rm II/I}(z)$]
 reduces as the redshift
 decreases, because massive halos appear abundant at low redshift and massive
 halos whose binding energy is much higher than before can reduce the feedback
 effect from SNe(see Eq.\ref{snfb} and Eq.\ref{snfb3}).

2. The radiative feedback mechanism can affect the SFR in both small
 ($T_{\rm vir}<10^4 K$) and large($T_{\rm vir}>10^4 K$) PopIII halos.
 In small halos, the shallow gravitational well and the poor
 cooling-efficiency cannot prevent the hot gas from escaping due to the
 strong photoevaporation from the massive OB stars.
 The radiative feedback can suppress the PopIII SFR considerably.
 In large halos, the ionizing background enhances the Jeans mass and
 makes the SFR suppressed more or less.

3. The SFRD in small PopIII halos($T_{\rm vir}<10^4 K$) is sensitive
 to $\tau_{\rm FB}$, which depends on $f^{\rm III}_{\rm esc}$
 and $n^{\rm III}_{\gamma}$ seriously. If $f^{\rm III}_{\rm esc}$
 and $n^{\rm III}_{\gamma}$ increases, $\tau_{\rm FB}$ will
 decreases and so does the SFRD. Furthermore, our results support
 this viewpoint: early star formation is likely to be self-regulated
 \cite{Ricotti02a,Ricotti02b,Yoshida03,cf05,KY05,Ricotti08}.

4. Although PopIII stars have high value of $f^{\rm III}_{\rm esc}$ and
 $n^{\rm III}_{\gamma}$, they are not able to reionize the universe considerably
 (about $20\%$ at $z\sim 14$). And the optical depth to Thomson scattering
 provided by PopIII stars is less than $22\%$.

In general, we find the radiative feedback is important to the formation of 
 the early generation stars. It suppresses the star formation considerably. 
 But the mechanical feedback from the SNe explosions is not able to affect
 the early star formation significantly. The radiative and mechanical
 feedback dominates the star formation rate of the PopII/I stars. 
 The feedback on SFRD from first generation stars is very
 strong and should not be neglected. However, their effect on the cosmic 
 reionization is not significant, which results in a small contribution to
 the optical depth of electrons $\tau_e$.

\section*{Acknowledgements} 
We thank Andrew Hopkins and Naveen Reddy for kindly providing the data on SFRD, Xu Kong
 for discussions, Li-Zhi Fang and M. Ricotti for comments.
 This work is partially supported by National Basic Research Program of China
 (2009CB824800), the National Natural Science Foundation
 (10733010,10673010,10573016), and the Program for New Century Excellent Talents
 in University.



\appendix
\section{How to get the SFR formula with SNe feedback}

Following Cen \& Ostriker (1992), a halo with mass $M$ has the initial baryonic
 gas $M_{\rm b}(0)=\f{\Omega_{\rm b}}{\Omega_{\rm m}}\times M$. 
 Later on, partial baryonic gas condenses into cold gas $M_{\rm cold}$.
It is reasonable to assume that the rates of the mass change
are proportional to $M_{\rm b}(t)$ at time $t$: 
\bea
\frac{\de M_{\rm b}(t)}{\de t}=-M_{\rm b}(t)\ \ \ {\rm and} \ \ \ 
\frac{\de M_{\rm cold}(t)}{de t}=+M_{\rm b}(t).
\label{a2}
\eea
This is also an implicit assumption in Cen \& Ostriker (1992). 
 By solving Eq.~(\ref{a2}) with the initial condition, we obtain
\bea
M_{\rm b}(t)=M_{\rm b}(0)\exp(-t)\ \ \ {\rm and} \ \ \ \nonumber \\
M_{\rm cold}(t)=\int\limits_{0}^{t} M_{\rm b}(t')\de t'=M_{\rm b}(0)
\left[1-\exp(-t)\right].
\label{a3}
\eea

Cold gas can form stars with efficiency $f_*$. And we assume that the newly formed
 star mass per unit time, $\dot{M}_{\rm SF}$, is proportional to the net mass of
 cold gas at that time $M^{\rm net}_{\rm cold}(t)$:
\bea
\dot{M}_{\rm SF}(t)=f_*\times M^{\rm net}_{\rm cold}(t),
\label{a4}
\eea here 
\bea
 M^{\rm net}_{\rm cold}(t)=M_{\rm cold}(t)-f_*^{-1} \int\limits_{0}^{t}
\dot{M}_{\rm SF}(t')\de t'.
\label{a5}
\eea
Solving Eq.~(\ref{a4}) with $\dot{M}_{\rm SF}(0)=0$
 (because $M^{\rm net}_{\rm cold}(0)=0$), we get
\bea
\frac{\de{M}_{\rm SF}(t)}{\de t}=f_* M_{\rm b}(0)\times t\exp(-t).
\label{a6}
\eea
If we replace $t$ with $\left[t(z)-t(z_{\rm c})\right]/
\left[\kappa t_{\rm dyn}(z_{\rm c})\right]$,
one can easily obtain Eq.~(\ref{SFR}). Since
 the feedback of supernovae explosions removes the cold gas at the rate\footnote{
In halos of mass $M\ge 10^8M_{\odot}$ or virial temperature of $T_{\rm vir}\ge
10^4K$ after $z=30$, even a PISN with $E_{\rm SN}=10^{53}ergs$ cannot make a
 substantial gas outflow\cite{KY05}. So we assume the baryons loss in a halo should not be
 taken into account.}
 $\dot M_{\rm II}^{\rm SN}=1.1 \epsilon_{\rm SN}^{\rm II}
 \dot{M}_{\rm SF}(t) \left(\frac{500 \hbox{km}\,
\hbox{s}^{-1}}{V_{\rm c}}\right)^2$; therefore,
\bea
 M^{\rm net}_{\rm cold}(t)=M_{\rm cold}(t)-f_*^{-1}\int\limits_{0}^{t}
\dot{M}_{\rm SF}(t')\de t'-\int\limits_{0}^{t}\dot M_{\rm II}^{\rm SN}(t')\de t'
\nonumber
\eea
\begin{equation}
=M_{\rm cold}(t)-\int\limits_{0}^{t}
\left[f_*^{-1}+1.1\epsilon_{\rm SN}^{\rm II}\left(\frac{500}
{V_{\rm c}}\right)^2\right]\dot{M}_{\rm SF}(t')\de t',
\label{a7}
\end{equation}
where the circular velocity $V_{\rm c}$ is a function with two variables $M$ and
 $z$ \cite{bl01}.
 Then $\left\{\f{\Omega_{\rm m}}{\Omega_{\rm m}(z)}\f{\Delta_c(z)}{18\pi^2}
\right\}^{1/3}$ varies between $0.65$ and $1.0$ when $z$ is between $30$ and zero.
 For simplicity we use its intermediate value $0.75$ in this work:

\bea
V_{\rm c}(M,z_{\rm c})=\sqrt{0.75}\times 23.4
\left(\f{M}{10^8h^{-1}M_{\odot}}\right)^{1/3}
\left(\f{1+z_{\rm c}}{10}\right)^{1/2}.
\label{a8}
\eea

Solving Eq.~(\ref{a4}) again with $\dot{M}_{\rm SF}(0)=0$, one can get the SFR with
 PopII/I SNe feedback:
\bea
\frac{\de M_{\rm SF}^{\rm II}(t)}{\de t}=\f{M_{\rm b}(0)}{S_{\rm II}(M,z_{\rm c})}
\Bigg\{\exp\big(-t\big)\nonumber \\
~-\exp\bigg\{-\Big[f^{\rm II}_*S_{\rm II}(M,z_{\rm c})+1\Big]t\bigg\}\Bigg\},
\label{a9}
\eea
where $S_{\rm II}(M,z_{\rm c})=1.82\times 10^9\,\epsilon_{\rm SN}^{\rm II}M^{-2/3}
(1+z_{\rm c})^{-1}$ for the feedback from PopII/I stars, while
$S_{\rm III}(M,z_{\rm c})=7.58\times 10^9\, \epsilon_{\rm SN}^{\rm III}
M^{-2/3}(1+z_{\rm c})^{-1}$ for the feedback from PopIII stars in halos with
$T_{\rm vir}>10^4 K$.
 After substituting $\left[t(z)-t(z_{\rm c})\right]/
\left[\kappa t_{\rm dyn}(z_{\rm c})\right]$ for $t$, we have the SFR with SNe feedback:

\bea
\dot M_{\rm SF}^{\rm II}(M,z,z_{\rm c})=
\f{\Omega_{\rm b} M}{\Omega_{\rm m}S_{\rm II}(M,z_{\rm c})\kappa t_{\rm dyn}}
\Bigg\{\exp\Big[-\f{t(z)-t(z_{\rm c})}{\kappa t_{\rm dyn}(z_{\rm c})}
\Big]\nonumber \\
-\exp\bigg\{-\Big[f^{\rm II}_*S_{\rm II}(M,z_{\rm c})+1\Big]\f{t(z)-t(z_{\rm c})}
{\kappa t_{\rm dyn}(z_{\rm c})}\bigg\}\Bigg\},\nonumber \\
\label{a10}
\eea
and 
\bea
\dot M_{\rm SF}^{\rm III}(M,z,z_{\rm c})=
\f{\Omega_{\rm b} M}{\Omega_{\rm m}S_{\rm III}(M,z_{\rm c})\kappa t_{\rm dyn}}
\Bigg\{\exp\Big[-\f{t(z)-t(z_{\rm c})}{\kappa t_{\rm dyn}(z_{\rm c})}
\Big]\nonumber \\
-\exp\bigg\{-\Big[f^{\rm III}_*S_{\rm III}(M,z_{\rm c})+1\Big]\f{t(z)-t(z_{\rm c})}
{\kappa t_{\rm dyn}(z_{\rm c})}\bigg\}\Bigg\}.
\nonumber \\
\label{a11}
\eea
 Obviously, Eq.~(\ref{a10}) and Eq.~(\ref{a11}) reduces to
 Eq.~(\ref{SFR}) when $S_{\rm III}(M,z_{\rm c})$ approaches to zero. 

\section{Radiative feedback from massive stars}

We focus on the radiative feedback from the
 ionizing background. 
 Similar to Samui et al. (2007), we describe the strength
 of radiative feedback via the circular velocity of halos
 $V_{\rm c}$ and the fraction of ionized hydrogen $Q_{\rm HII}$.
 {Considering the IGM in a unit volume after the ignition of the
 first stars, partial volume was reionized by the UV photons from
 massive stars in the massive halos($T_{\rm vir}>10^4 K$).}
 For the ionized part of the halo, their SFR will be suppressed
 by this ionizing background, which also depends on their mass. On the
 other hand, the SF will continue in the rest of the volume,
 where the IGM is neutral, so the total SFRD is contributed by
 these two parts at the same time:
\bea
\dot{\rho}_{\rm SF}(z,>10^4 K)=
\dot{\rho}^{\rm HI}_{\rm SF}(z)\left(1-Q_{\rm HII}(z)\right)
+\dot{\rho}^{\rm HII}_{\rm SF}(z)Q_{\rm HII}(z)\nonumber \\
=\Big[\dot{\rho}^{\rm HI}_{\rm II}(z)
+\dot{\rho}^{\rm HI}_{\rm III}(z,>10^4 K)\Big]
\left(1-Q_{\rm HII}(z)\right)\nonumber\\
+\Big[\dot{\rho}^{\rm HII}_{\rm II}(z)
+\dot{\rho}^{\rm HII}_{\rm III}(z,>10^4 K)\Big]Q_{\rm HII}(z)
.\nonumber\\
\label{b1}
\eea
where $\dot{\rho}^{\rm HI}_{\rm II}(z)$ stands for $\dot{\rho}^{\rm FB}_{\rm II}(z)$,
 $\dot{\rho}^{\rm HI}_{\rm III}(z)$ for $\dot{\rho}^{\rm FB}_{\rm III}(z,>10^4 K)$.

By combining Eq.~(\ref{dotng}), Eq.~(\ref{diffeq}) and Eq.~(\ref{b1}), one can get,

\bea
\f{\de Q_{\rm HII}(z)}{\de z}=\Theta(z)Q_{\rm HII}(z)+\Xi(z),
\label{b2}
\eea
here,
\bea
\Theta(z)=\f{1}{\bar{n}_{\rm H}^0}\Bigg\{n_{\gamma}^{\rm III}f_{\rm esc}^{\rm III}
\left(\f{\de {\rho}^{\rm HII}_{\rm III}(z,>10^4 K)}{\de z}-
 \f{\de {\rho}^{\rm HI}_{\rm III}(z,>10^4 K)}{\de z}\right)\nonumber \\
+n_{\gamma}^{\rm II}f_{\rm esc}^{\rm II}\left(\f{\de {\rho}^{\rm HII}_{\rm II}(z)}{\de z}-
 \f{\de {\rho}^{\rm HI}_{\rm II}(z)}{\de z}\right) \Bigg\}-
\alpha_{\rm B} (1+z)^3\bar{n}_{\rm H}^0 C(z)\f{\de t }{\de z} \nonumber,
\eea

\bea
\Xi(z)=\f{1}{\bar{n}_{\rm H}^0}\left\{n_{\gamma}^{\rm III}f_{\rm esc}^{\rm III}
\f{\de {\rho}^{\rm HI}_{\rm III}(z,>10^4 K)}{\de z}+n_{\gamma}^{\rm II}f_{\rm esc}^{\rm II}
\f{\de {\rho}^{\rm HI}_{\rm II}(z)}{\de z}  \right\} \nonumber.
\eea

\end{document}